\documentclass[proceedings]{JHEP}
 \usepackage{amsfonts}
 \usepackage{amssymb}
 \def\bsh{\backslash}
 \def\bdt{\dot \beta}
 \def\adt{\dot \alpha}

 \newfont{\bbbold}{msbm10 scaled \magstep1}
 \def\com{\mbox{\bbbold C}}

 \def\bbC{\mbox{\bbbold C}}

 \newfont{\goth}{eufm10 scaled \magstep1}

 \def\gg{\mbox{\goth g}}

 \def\gl{\mbox{\goth l}}

 \def\gp{\mbox{\goth p}}

 \def\gs{\mbox{\goth s}}

 \def\a{\alpha}
 \def\b{\beta}

 \def\vf{\varphi}

 \def\l{\lambda}

 \def\r{\rho}

 \def\th{\theta}
 
 \def\be{\begin{equation}}\def\ee{\end{equation}}
 \def\bea{\begin{eqnarray}}\def\eea{\end{eqnarray}}
 \def\ba{\begin{array}}\def\ea{\end{array}}

 \def\xz{\times}

 \def\bt{\bullet}
 \let\la=\label

 {}

 \def\bd{\begin{document}}
 \def\ed{\end{document}}
 \def\bea{\begin{eqnarray}}\def\barr{\begin{array}}\def\earr{\end{array}}
 \def\eea{\end{eqnarray}}
 \def\ft#1#2{{\textstyle{{\scriptstyle #1}\over {\scriptstyle #2}}}}
 \def\fft#1#2{{#1 \over #2}}
 \newcommand{\eq}[1]{(\ref{#1})}
 \def\eqs#1#2{(\ref{#1}-\ref{#2})}
 \def\det{{\rm det\,}}
 \def\tr{{\rm tr}}\def\Tr{{\rm Tr}}

\conference{Nonperturbative Quantum Effects 2000}

\title{Harmonic Superspaces and Superconformal Fields}
\author{Paul Heslop and Paul Howe.\\
The Maths Dept, King's College, The Strand, London, WC2R 2LS, England\\
E-mail:\email{pheslop@mth.kcl.ac.uk}} 

\abstract{Representations of four-dimensional superconformal groups on
 harmonic superfields are discussed. It is argued that any representation can
 be given as a superfield on many superflag
 manifolds. Representations
 on analytic superspaces do not require constraints. We discuss short
 representations and how to obtain them as explicit products of
 fundamental fields. We also discuss superfields that transform under
 supergroups.}
 
\begin{document}
 %%%%%%%%%%%%%%%%%%%%%%%%%%%%%%%%%%%%%%%%%%%%%%%%%%%%%%%%%%%%%%%%%%%%%%%%%%%%

 %%%%%%%%%%%%%%%%%%%%%%%%%%%%%%%%%%%%%%%%%%%%%%%%%%%%%%%%%%%%%%%%%%%%%%%%%%
 %%%%%
 \section{Introduction}

%%%%%%%%%%%%%%%%%%%%%%%%%%%%%%%%%%%%%%%%%%%%%%%%%%%%%%%%%%%%%%%%%%%%%%%%%%%%% 
The unitary irreducible representations of
 superconformal groups have assumed more
 significance recently in the light of the Maldacena conjecture \cite{mal}
 relating string theory or M-theory on
 $AdS \times S$ to superconformal field theories on
 the boundary. A particularly important class of operators that can
 arise consists of those operators which correspond to short representations of the
 superconformal group since these are expected to be protected from
 quantum corrections and thus not acquire anomalous dimensions
 \cite{fz}. Long representations are also of
 interest and are supposed to correspond to string states \cite{af2}.

 There are different methods of constructing these
 representations \cite{screp,gun}. In this talk however we discuss a way
 of constructing
 representations 
 explicitly as superfields using the method of
 para\-bolic induction, focusing on four dimensional superconformal
 groups, $SU(2,2|N)$.\footnote{Superconformal fields in harmonic
 superspaces have also been discussed recently in \cite{fs}.} This method was
 discussed in some detail in \cite{hh1}, although a direct comparison
 with the more algebraic group-theoretic
 results of \cite{screp,gun} was not made at the time. We complexify
 spacetime, 
and complexify the superconformal group to $SL(4|N)$ so that all the spaces of
interest become coset spaces of this group. 
 We claim that any representation may be given as a holomorphic
field on many
superflag spaces. On some spaces
 (e.g. super Minkowski space and chiral
 spaces) the
fields will require extra constraints, whereas on others (in
 particular analytic spaces) they will require no constraints, and this makes
tensoring representations together to produce other representations
straightforward.

In section 2 we briefly recall parabolic induction and illustrate
this in the bosonic context with the group $SL(N)$. In section 3 we
consider the full superconformal group $SL(4|N)$ and look at short
representations. Finally in section 4 we consider representations that
transform under supergroups, and give some specific examples.

This talk is based on \cite{HH} and work in progress.

%%%%%%%%%%%%%%%%%%%%%%%%%%%%%%%%%%%%%%%%%%%%%%%%%%%%%%%%%%%%%%%%%%%%%%%%%%%%%%
\section{The Bosonic case}

\subsection{Coset Spaces}
If $G$ is a Lie group and $P$ a subgroup, a coset space $M$ is the space
of (right) cosets: $M=P\bsh G$, and we obtain the fibre bundle:
$ G \rightarrow P\bsh G$,
with fibres P.

We can define a representation of $G$ on the space of equivariant maps
$F:G\rightarrow V$, where $V$ is the representation space of $P$, i.e. maps
 such that
\be
F(hu)=R(h)F(u) \label{F(hu)=}
\ee
where $u\in G$
 and $R$ is the representation of $P$ on $V$ (in practice these are
 fields with indices). 
The induced representation itself is given by $F\mapsto g\cdot F,\ g\in
G$ where
\be
(g\cdot F)(u)=F(ug). \label{g.F}
\ee

All the subgroups $P$ we are interested in will be parabolic subgroups
 (defined below), and the maps $F$ will be holomorphic maps. In this
 case the spaces are known as flag spaces.

 Let G be a complex, simple Lie group and let $\gg$ be its Lie
 algebra. 
For the case of $\gs\gl(N)$ we define the Borel
 subalgebra to be the algebra of all lower triangular matrices (with
 non-zero entries on the diagonal allowed), and a parabolic subalgebra
 $\gp$ is one which is block lower triangular:
 \be
 \begin{picture}(300,100)(-30,-40)
 \put(10,0){$\left(\ba{cccccccccc}
  \hspace{.2em} \bt \hspace{.2em}&\hspace{.2em} \bt \hspace{.2em}& &&& &&&&    \\
  \hspace{.2em} \bt \hspace{.2em}&\hspace{.2em} \bt \hspace{.2em}&&&&  &&&&\\
  \hspace{.2em} \bt \hspace{.2em}&\hspace{.2em} \bt \hspace{.2em}&\hspace{.2em} \bt \hspace{.2em}&\hspace{.2em} \bt \hspace{.2em}& & &&&&\\
  \hspace{.2em} \bt \hspace{.2em}&\hspace{.2em} \bt \hspace{.2em}&\hspace{.2em} \bt \hspace{.2em}&\hspace{.2em} \bt \hspace{.2em}& & &&&&\\
  \hspace{.2em} \bt \hspace{.2em}&\hspace{.2em} \bt \hspace{.2em}&\hspace{.2em} \bt \hspace{.2em}&\hspace{.2em} \bt \hspace{.2em}&\hspace{.2em} \bt \hspace{.2em}&\hspace{.2em} \bt \hspace{.2em}&\hspace{.2em} \bt \hspace{.2em}& &&\\
  \hspace{.2em} \bt \hspace{.2em}&\hspace{.2em} \bt \hspace{.2em}&\hspace{.2em} \bt \hspace{.2em}&\hspace{.2em} \bt \hspace{.2em}&\hspace{.2em} \bt \hspace{.2em}&\hspace{.2em} \bt \hspace{.2em}&\hspace{.2em} \bt \hspace{.2em}& &&\\
  \hspace{.2em} \bt \hspace{.2em}&\hspace{.2em} \bt \hspace{.2em}&\hspace{.2em} \bt \hspace{.2em}&\hspace{.2em} \bt \hspace{.2em}&\hspace{.2em} \bt \hspace{.2em}&\hspace{.2em} \bt \hspace{.2em}&\hspace{.2em} \bt \hspace{.2em}& &&\\
  &&&&&&&& .& \\
  &&&&&&&&&.
  \ea\right)$}
  \put(25,45) {$
 \left.\phantom{\ba{cc}
  \bullet&\bullet\\
  \bullet&\bullet\ea}\right\}
{\scriptscriptstyle k_1} $}
  \put(30,35) {$
 \left.\phantom{\ba{cccc}
  \bullet&\bullet&\bt&\bt\\
  \bullet&\bullet&\bt&\bt\\
  \bullet&\bullet&\bt&\bt\\
  \bullet&\bullet&\bt&\bt
 \ea}\right\}
{\scriptscriptstyle k_2} $}
  \put(40,15) {$
 \left.\phantom{\ba{ccccccc}
  \bullet&\bullet&\bt&\bt&\bt&\bt&\bt\\
 \bullet&\bullet&\bt&\bt&\bt&\bt&\bt\\
 \bullet&\bullet&\bt&\bt&\bt&\bt&\bt\\
 \bullet&\bullet&\bt&\bt&\bt&\bt&\bt\\
 \bullet&\bullet&\bt&\bt&\bt&\bt&\bt\\
 \bullet&\bullet&\bt&\bt&\bt&\bt&\bt\\
 \bullet&\bullet&\bt&\bt&\bt&\bt&\bt\ea}
  \right\}{\scriptscriptstyle k_3} $}
\end{picture}
 \ee
The corresponding Levi subalgebra is the block diagonal subalgebra,
i.e. $\gs(\gg \gl (k_1) \oplus \gg \gl (k_2-k_1) \oplus \ldots)$. A parabolic
 $\gp$ can also be represented by placing a cross on each of the nodes
 $k_1,k_2, \ldots, k_l$ of the Dynkin diagram for $\gs\gl(N)$ (see
 \cite{be}).

For example, complexified Minkowski space can be
 viewed as an open subset of the coset space $P\bsh SL(4)$, $SL(4)$
 being the complexified conformal group and $P$ the parabolic subgroup of
 matrices of the following shape:
 \be
 \left(\ba{cccc} \hspace{.2em} \bt \hspace{.2em}&\hspace{.2em} \bt \hspace{.2em}&&\\ \hspace{.2em} \bt \hspace{.2em}&\hspace{.2em} \bt \hspace{.2em}&&\\
 \hspace{.2em} \bt \hspace{.2em}&\hspace{.2em} \bt \hspace{.2em}&\hspace{.2em} \bt \hspace{.2em}&\hspace{.2em} \bt \hspace{.2em}\\ \hspace{.2em} \bt \hspace{.2em}&\hspace{.2em} \bt \hspace{.2em}&\hspace{.2em} \bt \hspace{.2em}&\hspace{.2em} \bt \hspace{.2em} \ea
 \right) 
\ee
where the bullets denote elements which do not have to be zero. The
 blank region can be thought of as corresponding to spacetime.  Indeed,
 we can choose a coset representative of the form
 \be
 M\ni x\mapsto s(x)= \left(\ba{cc} 1_2&x\\ 0_2&1_2 \ea \right)
\ee
 where each entry is a two-by-two matrix. From this one can easily
 work out the transformation of $x$ under the conformal group.
 The Levi subalgebra is $\gs(\gg \gl
 (2) \oplus \gg \gl(2))$ and the Dynkin diagram is
 \begin{picture}(30,10)(0,0) \put(0,0){\makebox[0pt][l]{$\bt
 \xz\bt$} \rule[.5ex]{1.8em}{.1ex} }
 \end {picture}.

%%%%%%%%%%%%%%%%%%%%%%%%%%%%%%%%%%%%%%%%%%%%%%%%%%%%%%%%%%%%%%%%%%%%
\subsection{Representations of $SL(N)$}
%%%%%%%%%%%%%%%%%%%%%%%%%%%%%%%%%%%%%%%%%%%%%%%%%%%%%%%%%%%%%%%%%%%%

Highest weight representations of $SL(N)$ can be specified by giving
$N-1$ integral Dynkin
labels, $a_i \geq 0$, $ i=1 \ldots N-1$ which are placed above the nodes of
the Dynkin diagram for $SL(N)$. Highest weight representations of
parabolic subgroups  are actually representations of the Levi subgroup
as the other bits act trivially. These can also be specified by giving
$N-1$ Dynkin labels, placed above
the corresponding Dynkin diagram with crosses through it (in this case
the labels above nodes with crosses through them can be negative, and
correspond to $\bbC$ charges, the remaining labels give the 
representations of the $\gs \gl(k_i)$ of the Levi algebra.) The
Borel-Weil theorem tells us that if we
pick a representation of $P$ with positive Dynkin labels, then
the induced representation of $G$ on holomorphic fields of $P\bsh G$
described above is isomorphic to the representation of $G$ with the
same Dynkin labels (see \cite{be}). Diagrammatically we have
\be
\ba{rcl}
&&\begin{picture}(150,20)
\put(0,0){\makebox[0pt][l]{$\bt\hspace{1.5em}\bt  \hspace{1.5em}\bt\hspace{1.5em}
\bt \hspace{1.5em}\bt    \hspace{1.5em}\bt\hspace{1.5em}\bt$} 
\rule[.5ex]{2em}{.1ex} 
$\hspace{0.3em} \cdots \hspace{.9em} 
\cdots \hspace{.9em} \cdots
\hspace{0.9em} \cdots \hspace{0.3em} $
\rule[.5ex]{2em}{.1ex} } 
\put(0,10){$\tiny  a_1
\hspace{1.5em} a_2\hspace{1.5em} a_{j} \hspace{1.8em} a_k \hspace{1.8em}
a_{l}\hspace{1em} a_{N-2}\hspace{.5em} a_{N-1}$}
\end {picture}\\
&\cong& 
\begin{picture}(150,30)
\put(0,0){\makebox[0pt][l]{$\bt\hspace{1.5em}\bt  \hspace{1.5em}\times
    \hspace{1.5em}
\bt \hspace{1.5em}\times \hspace{1.5em}\bt\hspace{1.5em}\bt$} 
\rule[.5ex]{2em}{.1ex} 
$\hspace{0.3em} \cdots \hspace{1.2em} 
\cdots \hspace{1em} \cdots
\hspace{1em} \cdots \hspace{0.3em} $
\rule[.5ex]{2em}{.1ex} } 
\put(0,10){$\tiny  a_1
\hspace{1.5em} a_2\hspace{1.5em} a_{j} \hspace{2em} a_k \hspace{2em}
a_{l}\hspace{1em} a_{N-2}\hspace{.5em} a_{N-1}$}
\end {picture} 
\ea
\ee
where crosses can be placed on any of the nodes of the right hand side. 

The simplest example of this formula
gives representations of $SL(2)$ as fields on $H\bsh SL(2) = \com
P^1$ where $H$ 
is the set of $2\times 2$ lower triangular matrices with unit
determinant. Diagrammatically we have
\be
\ba{c} p \\ \bt \ea \qquad \cong  \qquad
\ba{c} p \\ \times \ea
\ee
The right hand side of this equation represents the space of holomorphic tensor
fields of charge $p$ on $\com P^1$. This is a $p+1$-dimensional space
which is identified with the space of $p$th rank symmetric tensors
under $SL(2)$ (the left-hand side of the equation.)

%%%%%%%%%%%%%%%%%%%%%%%%%%%%%%%%%%%%%%%%%%%%%%%%%%%%%%%%%%%%%%%%%%%%%%%%%%%%%
\section{Superspaces}
%%%%%%%%%%%%%%%%%%%%%%%%%%%%%%%%%%%%%%%%%%%%%%%%%%%%%%%%%%%%%%%%%%%%%%%%%%%

We wish to extend the above formalism to the case of the
 superconformal group $SL(4|N)$.\footnote{We are eventually going to
 be interested in unitary representations of the real superconformal
 group $SU(2,2|N)$. However, homogeneous space techniques are more
 easily applied in the complex setting. One can return to real
 spacetime by taking $x$ real and $\varphi = \bar{\th}$.} Here we can
 have different, inequivalent Borel
 subgroups (see \cite{corn}). However, if we make a change of basis of
 $\com^{4|N}$ on 
 which the
 $SL(4|N)$ matrix acts, we can change an element in the Lie
algebra $\gs \gl(4|N)$ as follows:
\be
          g= \left( \ba{cc|c} \raisebox{-1.5ex}{\makebox[1em][l]{\hspace{.4em} 4}}&\phantom{4}& \\ \phantom{4}&\phantom{4}&\\ \hline &&N
                          \ea \right)
  \rightarrow
  \left( \ba{c|c|c}  2&& \\ \hline &N& \\ \hline &&2 \ea     \right). \label{eq:g}
  \ee
In this basis the particular choice of Borel subalgebra
 consistent with super Minkowski space consists of the lower
 triangular matrices, and\\ parabolic subalgebras consist of block
 lower triangular matrices, as in the bosonic case.

Super Dynkin diagrams can also be defined for $SL(4|N)$ as follows
\be
\begin{picture}(300,20)(-10,-10)
\put(0,0){\makebox[0pt][l]{$\bt\hspace{1.5em}\circ\hspace{1.5em}
    \underbrace{\bt \hspace{1.5em}
\bt\hspace{1.5em}\cdots \hspace{1.5em}\bt\hspace{1.5em}\bt}_{N-1}\hspace{1.5em}\circ\hspace{1.5em}
\bt$} 
\rule[.5ex]{6.45em}{.1ex} 
$\hspace{4.3em}$
\rule[.5ex]{6.6em}{.1ex}
 } 
\end {picture} 
\ee
In this diagram the $N-1$ central black nodes
represent the $\gs \gl(N)$ subalgebra. The two black nodes on the ends
represent
the space-time $\gs \gl(2)$ representations. The two white nodes are odd nodes
representing odd roots in the Lie algebra. This diagram is not a unique
diagram for $\gs \gl (4|N)$, but is the one which corresponds to the choice of
Borel
subalgebra above, and is thus the one which is consistent with super Minkowski
space. 
Different Borel subalgebras will lead to different Dynkin diagrams. Crosses can
be put anywhere on this diagram to represent parabolic
spaces as in the bosonic case.

 For example, complexified super Minkowski space has the form $P\bsh SL(4|N)$ where
 $P$ consists of matrices of the form
\be
 \left( \ba{cc|ccc|cc}
                        \hspace{.2em} \bt \hspace{.2em} & \hspace{.2em} \bt \hspace{.2em} &&&&&\\
                        \hspace{.2em} \bt \hspace{.2em} & \hspace{.2em} \bt \hspace{.2em} &&&&&\\
\hline                  \hspace{.2em} \bt \hspace{.2em} & \hspace{.2em} \bt \hspace{.2em} &\hspace{.2em} \bt \hspace{.2em} &. &\hspace{.2em} \bt \hspace{.2em} &&\\
                        . &.  &. &. &. &&\\
                        \hspace{.2em} \bt \hspace{.2em} & \hspace{.2em} \bt \hspace{.2em} &\hspace{.2em} \bt \hspace{.2em} &. &\hspace{.2em} \bt \hspace{.2em} &&\\
\hline                  \hspace{.2em} \bt \hspace{.2em} &\hspace{.2em} \bt \hspace{.2em} &\hspace{.2em} \bt \hspace{.2em} &.  &\hspace{.2em} \bt \hspace{.2em} &\hspace{.2em} \bt \hspace{.2em} &\hspace{.2em} \bt \hspace{.2em}\\
                        \hspace{.2em} \bt \hspace{.2em} &\hspace{.2em} \bt \hspace{.2em} &\hspace{.2em} \bt \hspace{.2em} &.  &\hspace{.2em} \bt \hspace{.2em} &\hspace{.2em} \bt \hspace{.2em} &\hspace{.2em} \bt \hspace{.2em} 
                        \ea \right)
\ee
and has corresponding Dynkin diagram
\be
\begin{picture}(300,0)(-5,0)
\put(0,0){\makebox[0pt][l]{$\bt\hspace{1.5em}\otimes\hspace{1.5em}\bt\hspace{1.5em}
\bt\hspace{1.5em}\cdots
\hspace{1.5em}\bt\hspace{1.5em}\bt\hspace{1.5em}\otimes \hspace{1.5em}
\bt$} 
\rule[.5ex]{6.7em}{.1ex} 
$\hspace{4.5em}$
\rule[.5ex]{6.9em}{.1ex}
 } 
\end {picture} 
\ee
The standard coset representative is
 \be
  M\ni z\mapsto s(z)=\left( \ba{c|c|c}
  1_2&\th &x  \\
 \hline
 0&1_N &\vf\\
  \hline
   0&0 &1_2\\
  \ea\right)
 \ee
 where $\vf$ denotes the $N$ dotted two-component spinorial coordinates
 which become
 the complex conjugates of the $\th$'s in the real case.

 We shall be interested in other superspaces which extend Minkowski
 space by an internal flag space. Such superspaces are called harmonic
 superspaces and were first introduced by GIKOS
 \cite{gikos}. Complexified $(N,p,q)$ harmonic superspace has
 the
 following
Dynkin diagram
\be
\begin{picture}(300,0)(-5,0)
\put(0,0){\makebox[0pt][l]{$\bt\hspace{1.5em}\otimes\hspace{1.5em}\bt\hspace{1.5em}
\times\hspace{1.5em}\bt
\hspace{1.5em}\times\hspace{1.5em}\bt\hspace{1.5em}\otimes \hspace{1.5em}
\bt$} 
\rule[.5ex]{4.5em}{.1ex} 
$\hspace{.4em} \cdots \hspace{1em} \cdots \hspace{1em} \cdots \hspace{1em}\cdots \hspace{.4em}$
\rule[.5ex]{4.5em}{.1ex}
 } 
\end {picture} 
\ee
where the middle crosses are on the $p$th and $(N-q)$th central
nodes. Locally, this space has the form of complex super Minkowski
space
times an internal flag space. The related
$(N,p,q)$ analytic superspace has the same body but fewer odd
coordinates.
It has the following Dynkin diagram 
\be
\begin{picture}(300,0)(-10,0)
\put(0,0){\makebox[0pt][l]{$\bt\hspace{1.5em}\circ\hspace{1.5em}\bt\hspace{1.5em}
\times\hspace{1.5em}\bt
\hspace{1.5em}\times\hspace{1.5em}\bt\hspace{1.5em}\circ \hspace{1.5em}
\bt$} 
\rule[.5ex]{4.5em}{.1ex} 
$\hspace{.2em} \cdots \hspace{1em} \cdots \hspace{1em} \cdots \hspace{1em}\cdots \hspace{.4em}$
\rule[.5ex]{4.5em}{.1ex}
 } 
\end {picture} 
\ee
This space has only $(N-p)$ $\th$'s and $(N-q)$ $\vf$'s. 
Generalised $(N,p,q)$ spaces can be defined, which have the same number of
$\th$'s and $\vf$'s as $(N,p,q)$ space, but have a different internal space.
These are given by the same Dynkin diagram as above, but with any number of
extra crosses inserted between the two already there.

%%%%%%%%%%%%%%%%%%%%%%%%%%%%%%%%%%%%%%%%%%%%%%%%%%%%%%%%%%%%%%%%%%%%%%%%%%%%
\subsection{Superconformal Representations}
%%%%%%%%%%%%%%%%%%%%%%%%%%%%%%%%%%%%%%%%%%%%%%%%%%%%%%%%%%%%%%%%%%%%%%%%%%%%%

Representations of the superconformal group \\ $SL(4|N)$ can be specified
by the following quantum numbers: Lorentz spin, $j_1, j_2$, dilation
weight, $L$, R-charge $R$, and the Dynkin labels of the internal
group, $a_1 \ldots a_{N-1}$. The unitary irreducible highest weight representations fall
in three series: A, B and C \cite{screp}. It is possible to define super-Dynkin
labels for the group $SL(4|N)$ as follows:
\be
 \begin{picture}(300,30)(-10,0)
\put(0,0){\makebox[0pt][l]{$\bt\hspace{1.5em}\circ\hspace{1.5em}\bt\hspace{1.5em}
\bt\hspace{1.5em}\cdots \hspace{1.5em}\bt\hspace{1.5em}\bt\hspace{1.5em}\circ\hspace{1.5em}
\bt$} 
\rule[.5ex]{6.45em}{.1ex} 
$\hspace{4.3em}$
\rule[.5ex]{6.6em}{.1ex}
 } 
\put(0,10){$\tiny  2j_1
\hspace{1em} S\hspace{2em} a_1 \hspace{2em} a_2 \hspace{5em}
a_{N-2}\hspace{1em} a_{N-1}\hspace{.5em} T \hspace{1.5em} 2j_2$}
\end {picture} 
\ee
where
\be \ba{rcl}
S&=&\frac12(L-R) +j_1 + \frac mN -m_1\\
T&=&\frac12(L+R) +j_2 -\frac mN \ea
\ee
and where $m_1 = \sum_{k=1}^{N-1} a_k$, $m=\sum_{k=1}^{N-1} k a_k $. Then
the three series correspond to the following conditions on the labels:
\be
\ba{lll}
A) \qquad & S \geq 2j_1 +1 \qquad& T \geq 2j_2 +1\\ \\
B)  & S \geq 2j_1 +1& T=j_2=0\\
or  & S= j_1 =0     &T \geq 2j_2 +1\\ \\
C)  & S= j_1 =0     &T=j_2=0   \ea
\ee
We are now in a position to apply the formalism of section 2 to the super case.
%%%%%%%%%%%%%%%%%%%%%%%%%%%%%%%%%%%%%%%%%%%%%%%%%%%%%%%%%%%%%%%%%%%%5
\subsection{Short Representations}
%%%%%%%%%%%%%%%%%%%%%%%%%%%%%%%%%%%%%%%%%%%%%%%%%%%%%%%%%%%%%%%%%%5

Short representations are characterised by
being short multiplets and thus having shorter range of spins than
unconstrained superfields on Minkowski superspace. Such
representations act naturally on superfields defined on analytic
superspaces since these have fewer odd coordinates than Min\-kowski
superspace.

The superfields on $(N,p,q)$ space should transform under
irreducible representations of Levi subalgebras of the form
$\gl=\gs(\gg\gl(2|p)\oplus\gg\gl(2|q)\oplus \gg\gl(r))$, $r=N-(p+q)$. However,
in order to ensure that the representations are indeed short these
superfields must not carry any spacetime indices. They must
therefore transform trivially under any supergroup factors of the
Levi subgroup.
 In the generic case this means that they transform
only under $\gs\gl(r)\oplus\com^2$.

In order to keep matters as simple as possible, we shall
concentrate for the time being on $(N,p,q)$ superspaces. The 
representations to
be studied can then be represented by modified Dynkin diagrams of
the following type:
\be
 \begin{picture}(300,30)(-10,0)
\put(0,0){\makebox[0pt][l]{$\bt\hspace{1.5em}\circ\hspace{1.5em}\bt\hspace{1.5em}
\times\hspace{1.5em}\bt\hspace{1.5em}\times\hspace{1.5em}\bt\hspace{1.5em}\circ\hspace{1.5em}
\bt$} 
\rule[.5ex]{4.3em}{.1ex} 
$\hspace{.3em} 
\cdots 
\hspace{.3em}$
\rule[.5ex]{4.7em}{.1ex}
$\hspace{.3em} 
\cdots 
\hspace{.3em}$
 \rule[.5ex]{4.3em}{.1ex} } 
\put(0,10){$\tiny  0
\hspace{2.5em} 0\hspace{2.5em} 0 \hspace{2.5em} a_p \hspace{2em}
a_i\hspace{1.5em} a_{N-q}\hspace{1em} 0 \hspace{2.5em} 0
\hspace{2.5em} 0$}
\end {picture} 
\ee
For the reasons discussed above the first $(p-1)$ and the last $(q-1)$ Dynkin labels must
vanish, leaving $(r-1)$ labels to specify the representation of the
central $\gs\gl(r)$ and two further labels which specify the
charges.

%%%%%%%%%%%%%%%%%%%%%%%%%%%%%%%%%%%%%%%%%%%%%%%%%%%%%%%%%%%%%%%%%%%%%%%%%%%%%%
\subsection{Massless Multiplets}
%%%%%%%%%%%%%%%%%%%%%%%%%%%%%%%%%%%%%%%%%%%%%%%%%%%%%%%%%%%%%%%%%%%%%%%%%%%%%%

Some simple examples of superconformal representations are given by on-shell 
massless multiplets, with maximal helicity s,  where $[\frac N2] \leq
2s <N$ ($[\frac N2]$ denotes the nearest integer greater than or equal
to $\frac N2$.) These have the following super-Dynkin labels:
\be
 \begin{picture}(300,30)(-10,0)
\put(0,0){\makebox[0pt][l]{$\bt\hspace{1.5em}\circ\hspace{1.5em}\bt\hspace{1.5em}
\bt\hspace{1.5em}\bt\hspace{1.5em}\bt\hspace{1.5em}\bt\hspace{1.5em}\circ\hspace{1.5em}
\bt$} 
\rule[.5ex]{4.3em}{.1ex} 
$\hspace{.3em} 
\cdots 
\hspace{.3em}$
\rule[.5ex]{4.3em}{.1ex}
$\hspace{.3em} 
\cdots 
\hspace{.3em}$
 \rule[.5ex]{4.3em}{.1ex} } 
\put(0,10){$\tiny  0
\hspace{2.5em} 0\hspace{2.5em} 0 \hspace{2.5em} 0 \hspace{2.5em}
1\hspace{2.5em} 0\hspace{2.5em} 0 \hspace{2.5em} 0
\hspace{2.5em} 0$}
\end {picture} 
\ee
where $a_p=1$ and all other Dynkin labels are 0 (here $p=2s$). These are described in (real) super Minkowski
space, $M$, by superfields $W$ which have $p$ totally
antisymmetric internal indices and which satisfy \cite{s,hst}
\begin{eqnarray}
\bar D_{\adt}^i W_{j_1 \ldots j_{p}} & = &
\frac{p(-1)^{p-1}}{N-p+1} \delta_{[j_1}^i \bar D_{\adt}^k W_{j_2 \ldots
j_{p}] k} \nonumber \\ D_{\alpha i}W_{j_1 \ldots j_{p}} & = &
D_{\alpha [ i} W_{j_1 \ldots j_{p}]} \label{DW}
\end{eqnarray}
For each such superfield there is a conjugate superfield $\tilde
W_{i_1\ldots i_{N-p}}$, and
these obey similar constraints.
When $s=\frac{1}{4}N$, the multiplet is self-conjugate.

We can extend this to $p=N$, since for such a superfield the
constraints \eq{DW} imply that it is anti-chiral. Its conjugate has
no indices and is chiral. Such a chiral field describes an on-shell
massless super multiplet (with maximum spin $N/2$) if it satisfies
the additional constraint
\be
D_{\a i} D^{\a}_j W=0.
\la{chiraleom}
\ee

The most natural spaces to put these representations on is $(N,p,N-p)$
analytic superspace, giving the Dynkin diagram
\be
 \begin{picture}(300,30)(-10,0)
\put(0,0){\makebox[0pt][l]{$\bt\hspace{1.5em}\circ\hspace{1.5em}\bt\hspace{1.5em}
\bt\hspace{1.5em}\times\hspace{1.5em}\bt\hspace{1.5em}\bt\hspace{1.5em}\circ\hspace{1.5em}
\bt$} 
\rule[.5ex]{4.3em}{.1ex} 
$\hspace{.3em} 
\cdots 
\hspace{.3em}$
\rule[.5ex]{4.3em}{.1ex}
$\hspace{.3em} 
\cdots 
\hspace{.3em}$
 \rule[.5ex]{4.3em}{.1ex} } 
\put(0,10){$\tiny  0
\hspace{2.5em} 0\hspace{2.5em} 0 \hspace{2.5em} 0 \hspace{2.5em}
1\hspace{2.5em} 0\hspace{2.5em} 0 \hspace{2.5em} 0
\hspace{2.5em} 0$}
\end {picture} 
\ee
On these spaces they don't satisfy any constraints (an example of this
will be given in the next section.)

There are many other ways of representing such multiplets which are
``less efficient'' in that the superspaces have more odd
coordinates. Following the discussion at the end of the previous
section we can simply place crosses where we like on the above
Dynkin diagram with the restriction, for the moment, that the cross furthest to the
left must be to the left of the node with the 1 above it, and the
cross furthest to the right must be to the right of this node, to avoid
super indices. As
we are talking about analytic superspaces, we do not want any crosses
on the end nodes, (these would correspond to super twistor spaces) . For example, any such field (excluding $p=0,N$) can
be realised on $(N,1,1)$ space:
\be
\begin{picture}(300,0)(-5,0)
\put(0,0){\makebox[0pt][l]{$\bt\hspace{1.5em}\circ\hspace{1.5em}\times\hspace{1.5em}
\bt \hspace{1.5em}\cdots \hspace{1.5em}\bt \hspace{1.5em}\times\hspace{1.5em}\circ\hspace{1.5em}
\bt$} 
\rule[.5ex]{7em}{.1ex} 
$\hspace{4.3em}$
\rule[.5ex]{7em}{.1ex}
 } 
\end {picture} 
\ee

To illustrate this procedure we consider the $N=4$ Maxwell
super multiplet which is represented on $N=4$ super Minkowski space
by the self-conjugate Sohnius superfield $W_{ij}$
\cite{sohn}. We can put this on $(4,2,2)$ analytic space \cite{hh2}
\be
 \begin{picture}(170,30)(-20,0)
\put(0,0){\makebox[0pt][l]{$\bt\hspace{1.5em}\circ\hspace{1.5em}\bt\hspace{1.5em}
\times\hspace{1.5em}\bt\hspace{1.5em}\circ\hspace{1.5em}
\bt$} 
\rule[.5ex]{13.5em}{.1ex} 
 } 
\put(0,10){$\tiny  0
\hspace{2.5em} 0\hspace{2.5em} 0 \hspace{3em} 
1\hspace{2.5em} 0\hspace{2.5em} 0 
\hspace{2.5em} 0$}
\end {picture} 
\ee
 where it
becomes a field with no indices, and a charge. In the Yang-Mills
 theory, we can multiply these
 together and get fields which correspond to the Kaluza-Klein states on
 $AdS^5$ in the $AdS/CFT$ correspondence. We could also, however
put it on $(4,1,1)$ analytic superspace \cite{bandos}
\be
 \begin{picture}(170,30)(-20,0)
\put(0,0){\makebox[0pt][l]{$\bt\hspace{1.5em}\circ\hspace{1.5em}\times\hspace{1.5em}
\bt\hspace{1.5em}\times\hspace{1.5em}\circ\hspace{1.5em}
\bt$} 
\rule[.5ex]{13.5em}{.1ex} 
 } 
\put(0,10){$\tiny  0
\hspace{2.5em} 0\hspace{2.5em} 0 \hspace{3em} 
1\hspace{2.5em} 0\hspace{2.5em} 0 
\hspace{2.5em} 0$}
\end {picture} 
\ee
on which it has an $SL(2)$
index: $W_{1r}, r \in \{2,3\}$. We can obtain all series C
representations with $R=0$ by
simply multiplying copies of this field, and taking irreducible
representations of $SL(2)$. Finally we could put it on $(4,1,0)$
analytic superspace 
\be
 \begin{picture}(170,30)(-20,0)
\put(0,0){\makebox[0pt][l]{$\bt\hspace{1.5em}\circ\hspace{1.5em}\times\hspace{1.5em}
\bt\hspace{1.5em}\bt\hspace{1.5em}\otimes\hspace{1.5em}
\bt$} 
\rule[.5ex]{13.8em}{.1ex} 
 } 
\put(0,10){$\tiny  0
\hspace{2.5em} 0\hspace{2.5em} 0 \hspace{3em} 
1\hspace{3em} 0\hspace{2.5em} 0 
\hspace{2.5em} 0$}
\end {picture} 
\ee
where it becomes a field with an $SL(3)$ index:
$W_{1r}, r\in \{2,3,4\}$. In this case, the field only obeys all the
constraints of the Sohnius field due to self-conjugacy. We can obtain all series C and B
representations with $R=0$ by multiplying copies of this field
together and taking irreducible representations under $SL(3)$. Doing
this using the Yang-Mills field and taking multiple traces corresponds
to BPS states in the $AdS/CFT$ correspondence \cite{fs}.

The above superfields have been defined on $(4,1,1)$ and $(4,1,0)$
superspaces which have the smallest possible internal
flags. It is possible to relax this, and use ``generalised'' $(N,p,q)$
spaces. For example we could use the
maximal flag space determined by the Borel subalgebra. In the
$(4,1,1)$ case for instance, this is the space
\be
\begin{picture}(170,30)(-20,0)
\put(0,0){\makebox[0pt][l]{$\bt\hspace{1.5em}\circ\hspace{1.5em}\times\hspace{1.5em}
\times\hspace{1.5em}\times\hspace{1.5em}\circ\hspace{1.5em}
\bt$} 
\rule[.5ex]{14em}{.1ex} 
 } 
\put(0,10){$\tiny  0
\hspace{2.5em} 0\hspace{2.5em} 0 \hspace{3em} 
1\hspace{3em} 0\hspace{2.5em} 0 
\hspace{2.5em} 0$}
\end {picture} 
\ee
We would then split the indices as $I=1,2,3,4$; this
enables us to define the field $W_{12}$. A disadvantage of this is that
multiplying such fields together to
get different representations is now no longer such a
simple procedure. 

%%%%%%%%%%%%%%%%%%%%%%%%%%%%%%%%%%%%%%%%%%%%%%%%%%%%%%%%%%%%%%
\section{Super Indices}
%%%%%%%%%%%%%%%%%%%%%%%%%%%%%%%%%%%%%%%%%%%%%%%%%%%%%%%%%%%%%%%

In the previous section we considered various short representations on
harmonic superspaces, and we insisted that they did not have any
super indices so we could explicitly see that the representations were
short. In this section we consider fields that do have
super indices. For simplicity we will restrict ourselves to $N=2$
analytic superspace. This has the Dynkin diagram
\be
\begin{picture}(180,10)
\put(20,0){\makebox[0pt][l]{$\bt\hspace{3em}\circ\hspace{3em}\times
    \hspace{2.7em}
\circ \hspace{3em}\bt$} \rule[.5ex]{3.5em}{.1ex}
\rule[.5ex]{7.1em}{.1ex}  \rule[.5ex]{3.5em}{.1ex}
 }
\end {picture}
\ee 
and corresponding parabolic subgroup of the form
\be
\left( \ba{ll} a^A{}_B & 0 \\ c_{A'B} & d_{A'}{}^{B'} \ea \right)
\ee
where each entry is a $(2|1) \times (2|1)$ matrix. The Levi subalgebra
(under which our fields transform) is $\gs \gl(2|1) \oplus \gs \gl(2|1)\oplus
\com$ (corresponding to the block diagonals), where the first $\gs
\gl(2|1)$ subalgebra  is carried by un-primed
indices, and the second by primed
indices. We may choose a local coset representative $s(X)$,
as follows:
\be
  s(X) = \left( \ba{cc} 1 & X \\ 0& 1 \ea \right) \in
SL(4|2)
\ee 
where each element is a $(2|1) \times (2|1)$ matrix, and 
\be 
X=\left( \ba{cc} \l^{\a} & x^{\a \adt} \\ y & \pi^{\adt} \ea \right).
\ee 
The important point here is that it takes two coordinate patches to
cover analytic superspace. If we denote these two sets by $U$ and $U'$
and put primes on the coordinates for $U'$ then in the overlap, the
two sets are related as follows:
\be
\ba{rcl} x' &= &x - \frac{\l \pi}{y},\\
         \l'&=&-\frac1y \l ,\\
         \pi'&=&\frac1y \pi ,\\
         y' &=& \frac 1y.  \ea
\ee
Requiring our fields to be holomorphic
on both patches puts restrictions on the fields, which are equivalent
to the constraints on Minkowski space.

We illustrate this first, briefly, with the example of the
hypermultiplet \cite{gikos}. This is the representation of $SL(4|2)$
with dilation
weight $L=1$, $R=0$, $j_1 = j_2=0$ but has a non-zero Dynkin
label $a_1=1$. It therefore has the following Dynkin diagram on
analytic superspace:
\be
\begin{picture}(180,20)
\put(20,0){\makebox[0pt][l]{$\bt\hspace{3em}\circ\hspace{3em}\times
    \hspace{2.7em}
\circ \hspace{3em}\bt$} \rule[.5ex]{3.5em}{.1ex}
\rule[.5ex]{7.1em}{.1ex}  \rule[.5ex]{3.5em}{.1ex}
 } 
\put(20,10){$\tiny  0
\hspace{4.8em} 0 \hspace{4.8em} 1 \hspace{4.8em} 0 \hspace{4.8em}
0$}
\end {picture}
\ee 
We read from this diagram that the field is invariant under both
$\gs \gl(2|1)$ subalgebras (i.e. our field has no super indices), but it does
have a $\com$ charge. Thus our field is specified by two local
holomorphic functions $W$ and $W'$ on $U$
and $U'$ respectively, such that in the 
overlap $U \cap U'$
\be
W(x,\l,\pi,y) = y W'(x',\l',\pi',y').
\ee
From this one can show that it has only a short expansion:
\be
\ba{rcl}
W(x,\l,\pi,y)& =& \vf_1(x) + y \vf_2(x)
+ \l^{\a} \psi_{\a}(x)\\& +&
\pi^{\adt}\chi_{\adt}(x) -\l^{\a}\pi^{\adt} \partial _{\a \adt} \vf_2
\ea 
\ee
with all the components satisfying their equations of motion. This is
the usual hypermultiplet with two complex scalar fields and two
complex Weyl fermions, all of which are physical and on-shell.

Consider next the $N=2$ on-shell Maxwell multiplet, which is usually a chiral field in super Minkowski space. This has dilation weight
1, R-charge -1 and all other quantum numbers 0. The Dynkin
diagram for this field on analytic superspace is
\be
\begin{picture}(180,20)
\put(20,0){\makebox[0pt][l]{$\bt\hspace{3em}\circ\hspace{3em}\times
    \hspace{2.7em}
\circ \hspace{3em}\bt$} \rule[.5ex]{3.5em}{.1ex}
\rule[.5ex]{7.1em}{.1ex}  \rule[.5ex]{3.5em}{.1ex}
 } 
\put(20,10){$\tiny  0
\hspace{4.8em} 1 \hspace{4.8em} 0 \hspace{4.8em} 0 \hspace{4.8em}
0$}
\end {picture}
\ee
 We can read off from the above
Dynkin diagram exactly how the fields transform under the Levi
 subalgebra. In particular, it
transforms non-triv\-ially under
the first $\gs \gl(2|1)$ superalgebra. In fact, we have a field
with one down-stairs un-primed super index. Again we have two local
holomorphic fields on $U$ and $U'$ which we denote $W_A = (W_{\a}, W)$
and $W'_A=(W'_{\a},W')$, and one can show that these are related as
follows in the intersection:
\bea
W_{\a} &=& y W'_{\a}\\
W      &=& W'-\frac1y \l^{\a}W_{\a}
\eea
giving us the following result:
\be
\ba{rcl}
W_{\a} &=& \r_{1\a} + y \r_{2\a} + \l^{\b}F_{\a \b} -
\pi^{\adt}\partial_{\a \adt}C  \\ 
&-&\l^{\b} \pi^{\bdt}\partial_{\b \bdt} \r_{2\a}\\
W    &=& C - \l^{\a} \r_{2 \a}
\ea
\ee 
again with all the components satisfying their equations of
motion. These components are all on-shell and we have reproduced the
$N=2$ Max\-well multiplet which is given on Minkowski space by a
chiral field satisfying the second order constraint (\ref{chiraleom}).

As a final example, consider the $N=2$ superconformal stress-energy multiplet. On super Minkowski space it is a scalar superfield $T$ satisfying
\be
D_{\a i} D_j^{\a} T =0.
\ee
It has Dilation weight 2, and all other quantum numbers are 0. On
analytic superspace it has the Dynkin diagram
\be
 \begin{picture}(170,20)(10,0)
\put(20,0){\makebox[0pt][l]{$\bt\hspace{3em}\circ\hspace{3em}\times
    \hspace{2.7em}
\circ \hspace{3em}\bt$} \rule[.5ex]{3.5em}{.1ex}
\rule[.5ex]{7.1em}{.1ex}  \rule[.5ex]{3.5em}{.1ex}
 } 
\put(20,10){$\tiny  0
\hspace{4.8em} 1 \hspace{4.8em} 0 \hspace{4.8em} 1 \hspace{4.8em}
0$}
\end {picture}
\ee 
and from this we see that it is given by the
superfield $T_{A'A}$. It has been explicitly checked that this does
indeed give the correct on-shell components.  This
representation can be obtained explicitly in two different ways on analytic superspace:
firstly by multiplying a Maxwell field and its conjugate together
\be
T_{A'A}= W_{A'} W_{A}
\ee
and secondly by multiplying two hypermultiplet fields together with a
derivative:
\be
T_{A'A}=W_1\partial_{A'A}W_2 -W_2\partial_{A'A}W_1.
\ee

%%%%%%%%%%%%%%%%%%%%%%%%%%%%%%%%%%%%%%%%%%
\section{Conclusion}
%%%%%%%%%%%%%%%%%%%%%%%%%%%%%%%%%%%%%%

In this talk we have shown how to obtain representations of the
superconformal group on certain coset spaces of this group. We claim
that any representation may be given as a tensor field on many different
superflag manifolds, if we allow the fields to transform under
supergroups.
On some spaces the fields may require constraints,
whereas on others, in particular analytic superspaces, no constraints
are required. This facilitates the tensoring together of different
representations. The super Dynkin diagrams provide a simple way of
giving all the information required for putting representations on
coset spaces. 

Superfield representations of the superconformal group are important
in the $AdS/CFT$ correspondence. For example, we believe it is possible
to obtain all $N=4$ superconformal representations explicitly by
multiplying copies of the
Maxwell superfield and applying derivatives on $(4,2,2)$
analytic superspace.
The formalism should be useful when considering correlation
functions in super Yang Mills theories \cite{anal}. There also appears to be a
similarity with the oscillator 
construction of superconformal representations \cite{gun}.

{\bf Acknowledgements.} P. Heslop thanks the TMR network and the
organisers of the conference for support.


\begin{thebibliography}{99}

\bibitem{mal}
J. Maldacena, {\sl The large N limit of superconformal field
theories and supergravity}, Adv. Theor. Math. Phys. {\bf 2} (1998)
231-252, hep-th/9711200

\bibitem{fz}
S. Ferrara and A. Zapparoni, {\sl Superconformal fields, multiplet shortening 
and the AdS$_5$/SCFT$_4$ correspondence} hep-th/9908163

\bibitem{af2}
L Andrianopoli and S Ferrara {On short and long $SU(2,2|4)$ multiplets
  in the AdS/CFT
correspondence}, Lett. Math. Phys. {\bf 48} (1999) 145,
hep-th/9812067.

\bibitem{screp}
M.Flato and C. Fronsdal,  Lett. Math. Phys. {\bf 8} (1984) 159;
V.K. Dobrev and V.B. Petkova, Phys. Lett. {\bf B162} (1985) 127,
Fortschr. Phys. {\bf 35} (1987) 537; B. Binegar, Phys. Rev. {\bf
D34} (1986) 525; B. Morel, A. Sciarrino and P. Sorba, Phys. Lett
{\bf B166} (1986) 69, erratum {\bf B167} (1986) 486.

\bibitem{gun}
M Gunaydin, D Minic, M Zagermann {\sl 4D Doubleton Conformal
  Theories, CPT and IIB String on $AdS_5 \times S^5$}, Nucl.Phys. B{\bf 534}
(1998) 96-120

\bibitem{fs}
S. Ferrara and E. Sokatchev, {\sl Short representations of
$SU(2,2|N)$ and harmonic superspace analyticity.} hep-th/9912168;
S. Ferrara and E. Sokatchev, {\sl Superconformal Interpretation of BPS
  States in AdS Geometries} hep-th/0005151

\bibitem{hh1}
 P.S. Howe and G.G. Hartwell, {\sl A superspace survey}, Class.
 Quant. Grav. {\bf 12} (1995) 1823-1880.

\bibitem{HH}
P. Heslop and P.S.Howe {\sl On Harmonic Superspaces and Superconformal
  Fields in Four Dimensions}, hep-th/0005135

\bibitem{be}
 R Baston and M Eastwood, {\it The Penrose Transform}, Oxford
 University Press 1989.

\bibitem{corn}
J.F.Cornwell {\sl Group Theory in Physics vol.III}, Academic Press
1989

\bibitem{gikos}
 A Galperin A, E Ivanov, S Kalitzin, V Ogievetsky and E Sokatchev,
 {\sl Unconstrained $N=2$ matter, Yang-Mills and supergravity
 theories in harmonic superspace}, Class. Quantum Grav. {\bf 1} (1984) 469

\bibitem{s}
W. Siegel, Nucl. Phys. {\bf B177} (1981) 325.

\bibitem{hst}
P.S. Howe, K.S. Stelle and P.K. Townsend, Nucl. Phys. {B181} (1981)
445; Nucl. Phys. {\bf B182} (1981) 332.

\bibitem{sohn}
M.F. Sohnius, Nucl. Phys. {\bf 136} (1978) 461.

\bibitem{hh2}
 G.G. Hartwell and P.S. Howe {\sl $(N,p,q)$ harmonic
 superspace}, Int J. Mod. Phys {\bf 10}, (1995) 3901-3919.

\bibitem{bandos}
I.A. Bandos {\sl Solution of linear equations in harmonic variables},
Theor.Math.Phys. {\bf 76} no. 2 (1988) 169-183. 

\bibitem{anal}
 P.S. Howe, E. Sokatchev and P.C. West, {\sl Three-point functions 
in N=4 Yang-Mills}, Phys. Lett. {\bf B444} (1998) 341-351, 
hep-th/9808162; B. Eden, P.S. Howe, C. Schubert, E. Sokatchev and
 P.C. West, {\sl 
Extremal correlators in four-dimensional SCFT}, Phys. Lett. {\bf 
B472} (2000) 323-331, hep-th/991015


\end{thebibliography}
\end{document}